# Elastic Bending Modulus of Monolayer Graphene

Qiang Lu,[1] Marino Arroyo,[2] and Rui Huang[1,*]

[1]Department of Aerospace Engineering and Engineering mechanics, University of Texas, Austin, TX 78712, USA
[2]Department of Applied Mathematics, LaCa`N, Universitat Polite`cnica de Catalunya (UPC), Barcelona 08034, Spain

A new formula for elastic bending modulus of monolayer graphene is derived analytically from an empirical potential for solid-state carbon-carbon bonds. Two physical origins are identified for the non-vanishing bending modulus of the atomically thin graphene sheet, one due to the bond angle effect and the other resulting from the bond order term associated with dihedral angles. The analytical prediction compares closely with *ab initio* energy calculations. Pure bending of graphene monolayers are simulated by a molecular mechanics approach, showing slight nonlinearity and anisotropy in the tangent bending modulus as the bending curvature increases. An intrinsic coupling between bending and in-plane strain is noted for graphene monolayers rolled into carbon nanotubes.

The unique two-dimensional (2D) lattice structure and physical properties of graphene has drawn tremendous interests recently. In particular, rippling of suspended graphene monolayers has been observed, with mesoscopic amplitude and wavelength [1]. Imaging of monolayer graphene sheets on silicon dioxide has also shown structural corrugation [2, 3]. Theoretical studies [4-6] have suggested that bending stiffness of monolayer graphene is critical in attaining the structural stability and morphology for both suspended and supported graphene sheets, which in turn could have important impacts on their electronic properties. Furthermore, single-walled carbon nanotubes are essentially graphene monolayers subjected to cylindrical bending along particular chiral directions. The energetics of carbon nanotubes is thus directly related to graphene through its bending property [7]. Apparently, different bending curvature (thus the tube radius) and bending directions have led to nanotubes of various mechanical and electronic properties [8].

While in-plane mechanical properties (elastic modulus and strength) of monolayer graphene have been deduced from experiments [9, 10], direct measurement of bending stiffness of monolayer graphene has not been reported in the literature. The often cited experimental value of 1.2 eV (or 0.192 nN-nm) was derived from the phonon spectrum of graphite [11]. Theoretically, bending modulus of monolayer graphene has been predicted based on empirical potentials [7,12-14] and *ab initio* calculations [15,16]. The fact that the atomically thin graphene monolayer has a finite bending modulus is in contrast with classical theories for plates and shells [17]. Bending modulus of an elastic thin plate scales with the cube of its thickness, namely, $D = dM/d\kappa \sim Eh^3$, where $M$ is the bending moment, $\kappa$ the bending curvature, $h$ the plate thickness, and $E$ the Young's modulus. The linear relationship between the classical bending modulus and the Young's modulus is a result of the classical Kirchhoff hypothesis for thin plates [17], which assumes linear variation of the strain and stress along the thickness of a plate. For a graphene monolayer, however, its physical thickness cannot be defined unambiguously [14], and the Kirchhoff hypothesis simply does not apply. Therefore, different physical origins must be sought for bending moment and bending modulus in graphene monolayers.

Based on the first-generation Brenner potential [18], a simple analytical form was derived for the bending modulus of monolayer graphene under infinitesimal bending curvature [13,14], namely

$$D_1 = \frac{\sqrt{3}}{2}\frac{\partial V_{ij}}{\partial \cos\theta_{ijk}} = \frac{1}{2}V_A(r_0)\frac{\partial b_{ij}}{\partial \theta_{ijk}}, \qquad (1)$$

where the interatomic potential takes the form

$$V_{ij} = V_R(r_{ij}) - \bar{b}V_A(r_{ij}), \qquad (2)$$

and $\bar{b} = (b_{ij}+b_{ji})/2$ is a function of the bond angles. At the ground state of graphene, the bond length $r_{ij} = r_0$ and the four bond angles associated with each bond are identical, i.e., $\theta_{ijk} = \theta_{jik} = 2\pi/3$ ($k \neq i, j$). Eq. (1) reveals that the physical origin of bending modulus comes from the bond angle effect in the interatomic potential. Consequently, any empirical potential with only the nearest neighbor (two-body) interactions would lead to zero bending modulus of the monolayer. The multibody interactions accounted for by the bond angles in the Brenner potential include second-nearest neighbors.

Using the second set of parameters for the Brenner potential [18], the bending modulus predicted by Eq. (1) is: $D_1 = 0.133$ nN-nm, or equivalently, 0.83 eV. This prediction however is considerably lower than that from *ab initio* energy calculations, $D = 3.9$ eV-Å$^2$/atom [16], or equivalently, 0.238 nN-nm (1.5 eV). Applying the same equation for the second-generation Brenner potential [19] leads to an even lower bending modulus: $D_1 = 0.110$ nN-nm (0.69 eV). The discrepancy suggests that the bond angle effect does not fully account for the bending stiffness of monolayer graphene. In this Letter, we demonstrate that, in addition to the bond angle effect, the effect of dihedral angles must be included in the consideration of bending energetics, which adds a significant contribution to the bending stiffness of monolayer graphene.

While many empirical potentials (including Tersoff [12] and first-generation Brenner potentials [18]) consider multibody interactions up to the second-nearest neighbors, the second-generation Brenner potential includes the third nearest neighbors via a bond order term associated with dihedral angles [19]. Taking the same general form as Eq. (2), the bond order function of the second-generation Brenner potential is written as

$$\overline{b} = (b_{ij}^{\sigma-\pi} + b_{ji}^{\sigma-\pi})/2 + b_{ij}^{DH} + \Pi_{ij}^{RC}, \qquad (3)$$

Where $(b_{ij}^{\sigma-\pi} + b_{ji}^{\sigma-\pi})/2$ is a function of bond angles similar to that in the first-generation Brenner potential, $b_{ij}^{DH}$ is a function of dihedral angles, and $\Pi_{ij}^{RC}$ represents the influence of radical energetics and $\pi$-bond conjugation. For a perfect graphene lattice with no vacancy, $\Pi_{ij}^{RC} = 0$, and

$$b_{ij}^{DH} = \frac{T_0}{2} \sum_{k,l(\neq i,j)} \left[ \left(1 - \cos^2 \Theta_{ijkl}\right) f_c(r_{ik}) f_c(r_{jl}) \right], \qquad (4)$$

where $T_0 = -0.00809675$ and $f_c(r)$ is a cut-off function so that only the nearest neighbors ($k$ and $l$) of the atoms $i$ and $j$ are considered in the calculations of the dihedral angles. The dihedral angle is defined as

$$\cos \Theta_{ijkl} = \mathbf{n}_{jik} \cdot \mathbf{n}_{ijl}, \qquad (5)$$

where $\mathbf{n}_{jik}$ and $\mathbf{n}_{ijl}$ are the unit normal vectors to the planes of the triangles $jik$ and $ijl$, respectively. Each C-C bond in the graphene lattice is associated with four dihedral angles. At the ground state of graphene, the dihedral angles are either $0$ or $\pi$, and thus $b_{ij}^{DH} = 0$. However, the dihedral term becomes nonzero upon bending of the graphene monolayer.

The bending modulus of graphene is derived from the strain energy function $W$ as $D = \partial^2 W / \partial \kappa^2$, where $W$ depends on the bond lengths, bond angles, and dihedral angles in a unit cell of graphene and can be obtained by summing up the interatomic potential energy in the unit cell, namely

$$W = \frac{1}{S_0} \sum_{i,j,k,l} V_{ij}(r_{ij}, \theta_{ijk}, \Theta_{ijkl}), \qquad (6)$$

and $S_0 = 3\sqrt{3}r_0^2/2$ is the area of the unit cell at the ground state. It can be shown that, at the ground state of graphene,

$$\frac{\partial r_{ij}}{\partial \kappa} = 0, \quad \frac{\partial \theta_{ijk}}{\partial \kappa} = 0, \quad \frac{\partial}{\partial \kappa}(\cos \Theta_{ijkl}) = 0. \qquad (7)$$

and

$$\frac{\partial W}{\partial r_{ij}} = 0, \quad \frac{\partial W}{\partial \theta_{ijk}} = -\frac{2V_A(r_0)}{S_0} \frac{\partial b_{ij}^{\sigma-\pi}}{\partial \theta_{ijk}},$$

$$\frac{\partial W}{\partial \cos \Theta_{ijkl}} = \frac{T_0}{S_0} V_A(r_0) \cos \Theta_{ijkl}. \qquad (8)$$

Thus, the bending modulus under infinitesimal bending curvature from the ground state is

$$D = -\frac{2V_A(r_0)}{S_0} \frac{\partial b_{ij}^{\sigma-\pi}}{\partial \theta_{ijk}} \sum_{i,j,k} \frac{\partial^2 \theta_{ijk}}{\partial \kappa^2}$$
$$+ \frac{T_0 V_A(r_0)}{S_0} \sum_{i,j,k,l} \left( \cos \Theta_{ijkl} \frac{\partial^2 \cos \Theta_{ijkl}}{\partial \kappa^2} \right) \qquad (9)$$

As shown in Ref. 13, $\sum_{i,j,k} \frac{\partial^2 \theta_{ijk}}{\partial \kappa^2} = \frac{-9}{8\sqrt{3}} r_0^2$. Using a method of asymptotic expansion, we have shown that

$$\sum_{i,j,k,l} \left( \cos \Theta_{ijkl} \frac{\partial^2 \cos \Theta_{ijkl}}{\partial \kappa^2} \right) = -\frac{21 r_0^2}{2}. \qquad (10)$$

Finally, we obtain a new form for the bending modulus of monolayer graphene:

$$D = \frac{V_A(r_0)}{2} \left( \frac{\partial b_{ij}^{\sigma-\pi}}{\partial \theta_{ijk}} - \frac{14 T_0}{\sqrt{3}} \right). \qquad (11)$$

While the first term on the right-hand side of Eq. (11) is identical to Eq. (1), the second term results from the effect of dihedral angles in the second-generation Brenner potential. With the additional term, Eq. (11) predicts that $D = 0.225$ nN-nm (1.4 eV), very close to the *ab initio* calculations [16]. It is thus suggested that the effect of dihedral angles plays an important role in bending of graphene monolayers. In general, the multibody interactions including both the second and third nearest neighbors in an atomic monolayer result in a finite bending stiffness of the atomically thin membrane.

To validate the analytical form for the bending modulus of monolayer graphene, we carry out atomistic simulations in which graphene monolayers are rolled into cylindrical tubes. The static molecular mechanics (MM) approach is adopted to calculate the strain energy, from which the bending moment and bending modulus can be deduced. In previous studies [12,13,16], strain energy of fully relaxed carbon nanotubes was calculated as a function of the curvature ($\kappa = 1/R$). We found that, relative to the ground state of graphene, the deformation of a fully relaxed carbon nanotube involves both the bending curvature and in-plane strain [20]. Consequently, the strain energy density of fully relaxed carbon nanotubes includes contributions from in-plane strain and thus cannot be written simply as a quadratic function of the curvature alone. To achieve pure bending (with zero in-plane strain) of graphene monolayers, in our simulations the total potential energy is minimized under the constraint that the tube radius and length do not change. The constraint on the tube length is easily applied by the periodic boundary condition along the axial direction. To enforce the constraint on the tube radius, the graphene is first rolled up by mapping a 2D sheet (width = $a$) onto a cylindrical tube (radius $R = a/2\pi$). Next, the potential energy is minimized by internal relaxation between the two sublattices of graphene, with one sublattice fixed and the other allowed to relax. In this way, the overall tube radius does not change. The resulting tubes from these simulations are not fully relaxed; in other words, external reaction forces are required to keep the tube dimensions from relaxing, in both the axial and radial directions. With a pure bending deformation, the strain energy density of the constrained tube depends only on the tube radius or curvature. In the linear elastic regime, we have $W = D\kappa^2/2$, and the bending moment is simply, $M = dW/d\kappa = D\kappa$.

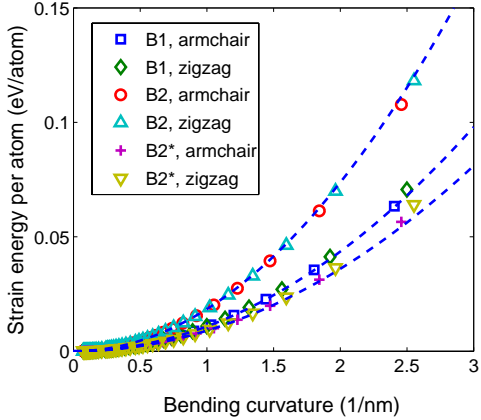

Fig. 1. Strain energy per atom as a function of curvature for pure bending of monolayer graphene along armchair and zigzag directions, obtained from different empirical potentials. The quadratic function, $W = D\kappa^2/2$, is plotted as the dashed line using the analytical bending modulus for each potential.

Fig. 1 plots the strain energy for pure bending of graphene monolayers along the armchair and zigzag directions. For comparison, the results from both the first and second-generation Brenner potentials (B1 and B2) are shown. To further highlight the effect of dihedral angles, also shown are the results from simulations ignoring the dihedral term in the second-generation Brenner potential (B2*). Clearly, the strain energy for B2 is systematically higher than the other two. The dihedral term contribute significantly to the bending energy. The corresponding bending moments are obtained by numerically differentiating the strain energy with respect to the bending curvature, as plotted in Fig. 2. For all three potentials, the bending moment increases almost linearly with the curvature up to 2 nm$^{-1}$, with slight nonlinearity at large curvatures. By further differentiating the bending moment with respect to the curvature, we obtain tangent bending modulus, as plotted in Fig. 3. The tangent modulus at small curvatures agrees closely with the analytical prediction. At larger curvatures (nanotubes of small radii), the tangent bending modulus deviates slightly as a result of nonlinearity. Due to the effect of dihedral angles, the tangent modulus obtained from the potential B2 is considerably higher than those from B1 and B2*.

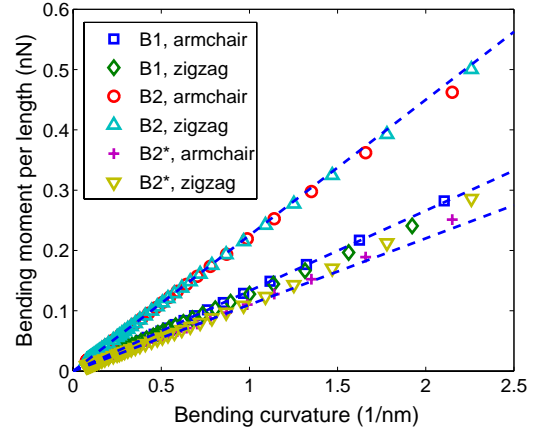

Fig. 2. Bending moment versus bending curvature of graphene along armchair and zigzag directions. The linear elastic bending moment-curvature relation, $M = D\kappa$, is plotted as the dashed line using the analytical bending modulus.

Fig. 3 shows that the tangent bending moduli along the zigzag and armchair directions are essentially identical at small curvatures, but become increasingly different as the curvature increases. As expected, the monolayer graphene at the ground state is elastically isotropic due to the hexagonal symmetry of graphene lattice, and the bending modulus at the linear elastic regime as predicted by Eq. (11) is independent of the bending direction. However, the lattice symmetry is distorted by the bending deformation and the monolayer graphene becomes slightly anisotropic at the nonlinear regime. Stronger anisotropy has been noted for the in-plane elastic moduli of monolayer graphene under finite stretch [20].

It is noted that, due to the constraint on the tube radius and length, the strain energy of pure bending (Fig. 1) is slightly higher than the corresponding strain energy in fully relaxed carbon nanotubes [13]. To illustrate the effect of the constraint, Fig. 4 plots the strain energy as a function of the tube radius for a (10, 0) carbon nanotube. The tube radius is gradually increased in the MM calculations, while the tube length remains fixed. Only the second-generation Brenner potential with the dihedral term is used here. As shown in Fig. 4, the strain energy reaches a minimum at $R/R_0 \sim 1.013$, where $R_0 = 0.397$ nm is the tube radius before relaxation. The minimum energy is a few percent lower than the pure bending energy, and it compares closely with the corresponding strain energy by MM calculations without imposing any constraint on the tube radius. Therefore, relaxation of the radial constraint alone leads to increase of the tube radius by about 1.3%, or an in-plane strain $\varepsilon = 0.013$ in the circumferential direction of the tube. Further relaxation in the axial direction would result in even lower strain energy along with slightly different tube radius and length for a fully relaxed nanotube.

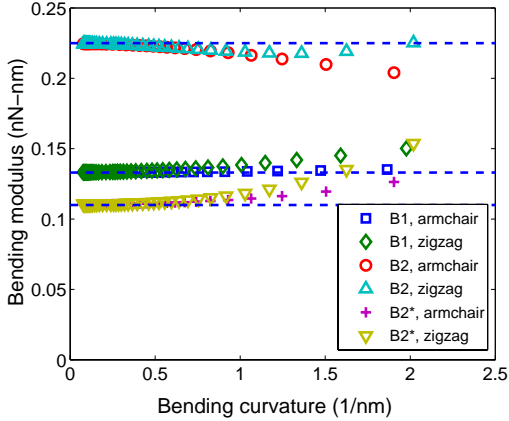

Fig. 3. Tangent bending modulus of monolayer graphene as a function of bending curvature along armchair and zigzag directions. The analytical result for the linear elastic bending modulus (independent of curvature) is plotted as the dashed line for each potential.

To understand the energy reduction and radius increase in the relaxed tube, one may consider the total strain energy as the sum of the bending energy and the in-plane strain energy. As the tube radius increases, the bending energy ($W_B = D\kappa^2/2$) decreases and the in-plane strain energy ($W_M = C\varepsilon^2/2$) increases, where $D$ and $C$ are the elastic moduli for bending and in-plane stretch, respectively. However, with $D = 0.225$ nN-nm and $C = 289$ N/m for the second-generation Brenner potential [13,20], the competition leads to a minimum energy at $\varepsilon \sim 0.005$ for the (10,0) nanotube, much smaller than that in Fig. 4. More accurately, we expand the strain energy with respect to the bending curvature and membrane strain, to the leading orders,

$$W(\kappa,\varepsilon) \approx \frac{1}{2}D\kappa_0^2 + M(\kappa - \kappa_0) + \sigma\varepsilon + \frac{1}{2}D(\kappa - \kappa_0)^2 + \frac{1}{2}C\varepsilon^2,$$
(12)

where $\kappa_0 = 1/R_0$ is the curvature before relaxation, $M = (\partial W/\partial \kappa)_{\varepsilon=0}$ is the bending moment, and $\sigma = (\partial W/\partial \varepsilon)_{\varepsilon=0}$ is the in-plane membrane force in the circumferential direction for the un-relaxed tube. The strain energy in Eq. (12) has a minimum at $\varepsilon = -(\sigma - D\kappa_0^2)/(C + 3D\kappa_0^2)$. For the (10,0) nanotube, we find that $\sigma = -2.38$ N/m and the agreement between Eq. (12) and the MM calculations (Fig. 4) is excellent up to a few percent of the in-plane strain. Therefore, the tube is subject to a compressive membrane force along the circumferential direction before relaxation. The compressive membrane force may be qualitatively understood as a result of the shortening of the bond lengths in the un-relaxed tube, relative to the bond length at the ground state of graphene. The imposed constraint over the tube radius effectively applies an external pressure onto the tube, balancing the internal membrane force. The external pressure can be obtained from the Laplace-Young equation, namely, $p = \sigma/R_0 = 6.0$ GPa. The deduced pressure compares closely with *ab initio* calculations of carbon nanotubes under hydrostatic pressure [21]. We note that the induced membrane force is in clear contrast with the classical plate theory which predicts zero membrane force under the pure bending condition [17]. This suggests an intrinsic coupling between bending and in-plane strain due to the discrete nature of the graphene lattice.

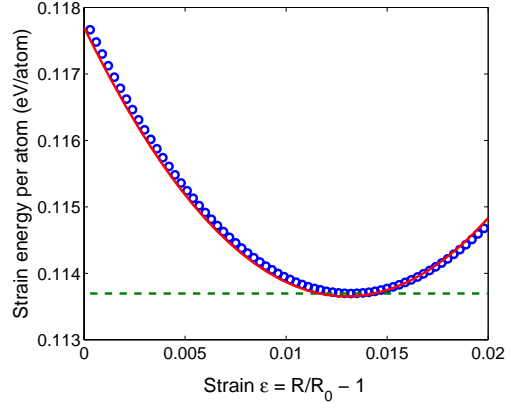

Fig. 4. Relaxation of strain energy for a (10, 0) carbon nanotube. The atomistic calculations are shown by open circles and the prediction by Eq. (12) as the solid curve. The dashed line indicates the strain energy from an atomistic simulation without imposing any constraint on the tube radius.

In summary, we have shown that the effects of bond angles (second nearest neighbors) and dihedral angles (third nearest neighbors) in the empirical interatomic potential are both physical origins for the non-vanishing bending modulus of the atomically thin graphene monolayers. A new analytical expression for the elastic bending modulus is derived, which compares closely with *ab initio* calculations. Slight nonlinearity and anisotropy are noted for tangent bending modulus calculated from molecular mechanics based atomistic simulations. An intrinsic coupling between bending and in-plane strain is suggested, which leads to a compressive membrane force under pure bending and reduction of the strain energy in relaxed carbon nanotubes.

The authors (RH and QL) gratefully acknowledge financial support by the US Department of Energy through Grant No. DE-FG02-05ER46230.

*Corresponding author: ruihuang@mail.utexas.edu.